\documentclass{appolb}
\usepackage{epsfig}

\begin{document}
\title{Testing BFKL evolution with Mueller-Navelet jets%
\thanks{Presented at the International Symposium on Multiparticle Dynamics, ISMD07}%
}
\author{Cyrille Marquet
\address{RIKEN Brookhaven Research Center, BNL, Upton, NY 11973, USA}
}
\maketitle
\begin{abstract}

We study the correlation in azimuthal angle between Mueller-Navelet jets produced in hadron-hadron collisions. We argue that this observable would test the BFKL approach at next-to-leading logarithmic accuracy. In order to motivate such a measurement, we give predictions for jets separated by large rapidity intervals within the Tevatron and LHC ranges.

\end{abstract}
  
\section{Motivation}

Mueller-Navelet jets \cite{mnjets} in hadron-hadron scattering are two jets produced in each of the forward directions. In standard perturbative QCD calculations, the partonic cross section is computed at fixed order with respect to the strong coupling $\alpha_s,$ while the large logarithms coming from the strong ordering between the hadronic scales and the jets transverse momenta are resummed using the DGLAP evolution equation for the parton densities. In the high-energy regime, in which the jets are separated by a large rapidity interval, other large logarithms arise in the hard cross section itself, due to the strong ordering between the total energy and the hard scales. These can be resummed using the BFKL equation, up to next-leading (NLL) logarithmic accuracy \cite{nllbfkl}. 

Although the NLL corrections are accompanied by spurious singularities due to the truncation of the BFKL expansion at a fixed order, it was realised that renormalisation-group constrained regularisations can cure the problem and lead to reasonable NLL-BFKL kernels. While phenomenological studies still require approximations, strong hints of NLL-BFKL effects have been observed in forward-jet production in lepton-hadron collisions \cite{nllfj}.
The present study is devoted to the correlation in azimuthal angle between Mueller-Navelet jets in hadron-hadron collisions. The goal is to motivate this feasible measurement at the Tevatron (Run 2) and at the LHC.

\section{The azimuthal correlation between Mueller-Navelet jets}

We denote $\sqrt{s}$ the total energy of the collision, $k_1$ and $k_2$ the transverse momenta of the two forward jets and $y_1\!>\!0$ and $y_2\!<\!0$ their rapidities. $\Delta\Phi\!=\!\pi\!-\!\phi_1\!+\!\phi_2$ measures the relative azimuthal angle between the two jets, as $\phi_1$ and
$\phi_2$ are the jets angles in the plane transerve to the collision axis. We shall work with the following kinematic variables:
\begin{equation}
\Delta\eta\!=\!y_1\!-\!y_2\ ,\hspace{0.5cm}y\!=\!\frac{y_1\!+\!y_2}2\ ,\hspace{0.5cm}
Q\!=\!\sqrt{k_1k_2}\ ,\hspace{0.5cm}\hspace{0.5cm}\!R=\!\frac{k_2}{k_1}\ .
\end{equation}

We are interested in an observable that is suitable to study the azimuthal decorrelation of the jets as a function of their rapidity separation $\Delta\eta$ and of the ratio of their transverse momenta $R:$
\begin{equation}
2\pi\left.\frac{d\sigma}{d\Delta\eta dR d\Delta\Phi}\right/\frac{d\sigma}{d\Delta\eta dR}=
1+\frac{2}{\sigma_0(\Delta\eta,R)}\sum_{p=1}^\infty \sigma_p(\Delta\eta,R) 
\cos(p\Delta\Phi)\ .\label{obs}
\end{equation}
We have expressed the normalized cross-section (\ref{obs}) in terms of the Fourier coefficients
$\sigma_p(\Delta\eta,R)$ given by
\begin{eqnarray}
\sigma_p(\Delta\eta,R)=\int_{E_T}^\infty \frac{dQ}{Q^3}
\left\{\int_{y_<}^{y_>} dy\ x_1f_{eff}(x_1,Q^2/R)x_2f_{eff}(x_2,Q^2R)\right\}
\nonumber\\\alpha_s(Q^2/R)\alpha_s(Q^2R)G(Q,R,Y)
\label{coeff}\end{eqnarray}
where $x_1$ and $x_2$ denote the longitudinal fraction of momentum of the jets with respect to the incident hadrons. In the high-energy regime in which the rapidity interval between the two jets is assumed to be very large, they are given by $x_1\!=\!k_1e^{y_1}/\sqrt{s}$ and $x_2=k_2e^{-y_2}/\sqrt{s}.$

The effective parton distribution $f_{eff}(x,k^2)$ resums the large logarithms
$\log(k^2/\Lambda_{QCD}^2).$ It is given by $f_{eff}\!=\!g\!+\!C_F(q\!+\!\bar{q})/N_c$ in terms of $g$ (resp. $q$, $\bar{q}$), the gluon (resp. quark, antiquark) distribution in the incident hadron. Since the Mueller-Navelet jet measurement involves perturbative values of $k_1$ and $k_2$ and moderate values of $x_1$ and $x_2,$ the cross-section features the collinear factorization of
$f_{eff},$ with $k_1^2$ and $k_2^2$ chosen as factorization scales.

In (\ref{coeff}), we choose to apply the rapidity cuts $y_>\!=\!-y_<\!=\!0.5$ to enforce the symmetric situation $y_2\!\sim\!-y_1.$ For the transverse momentum cut $E_T,$ we will consider two options: $E_T\!=\!20\ \mbox{GeV}$ for the Tevatron (Run 2) and $E_T\!=\!50\ \mbox{GeV}$ for the LHC. We recall that the respective center-of-mass energies are $\sqrt{s}\!=\!1960\ \mbox{GeV}$ and
$\sqrt{s}\!=\!14\ \mbox{TeV}.$

The BFKL Green function $G(Q,R,Y)$ is the main object of the study, it resums the large logarithms $\log(s/Q^2).$ At NLL accuracy, there is an ambiguity corresponding to the specific regularisation procedure. The results displayed in the following are obtained with the S4 scheme of \cite{salam}.

\section{The BFKL Green function at NLL accuracy}

There are many complications when performing a full NLL calculation in the BFKL framework. 
In fact, in formula (\ref{coeff}), we have already made the approximation to use leading-order impact factors. Dealing with next-to-leading order impact factors is possible, but goes beyond the scope of our study. The NLL-BFKL Green function is given by
\begin{equation}
G(Q,R,\Delta\eta)=
\int_{1/2-\infty}^{1/2+\infty}\frac{d\gamma}{2i\pi}R^{-2\gamma}
\ e^{\bar\alpha(Q^2)\chi_{eff}[p,\gamma,\bar\alpha(Q^2)]\Delta\eta}
\end{equation}
with the QCD running coupling $\bar\alpha(k^2)\!=\!\alpha_s(k^2)N_c/\pi
\!=\!\left[b\log\left(k^2/\Lambda_{QCD}^2\right)\right]^{-1}$ where $b=11/12\!-\!N_f/6N_c.$

The NLL-BFKL effects are phenomenologically taken into account by the 
effective kernel $\chi_{eff}(p,\gamma,\bar\alpha).$ For each value of $p,$ the
NLL-BFKL kernels provided by the regularisation procedure $\chi_{NLL}(p,\gamma,\omega)$ 
depend on $\gamma,$ the Mellin variable conjugate to $R^2$ and $\omega,$ the 
Mellin variable conjugate to $s/Q^2.$ In each case, the NLL kernels obey a consistency condition \cite{salam} which allows to reformulate the problem in terms of the effective kernel $\chi_{eff}(p,\gamma,\bar\alpha):$ it is obtained from the NLL kernel $\chi_{NLL}(p,\gamma,\omega)$ by solving the implicit equation
\begin{equation}
\chi_{eff}=\chi_{NLL}\left(p,\gamma,\bar\alpha\ \chi_{eff}\right)\ .
\end{equation}
The NLL-BFKL kernel $\chi_{NLL}(p,\gamma,\omega)$ is given in \cite{nllmnj} (see also
\cite{sabsch}), where the regularisation procedure of \cite{salam} is extended to nonzero conformal spins $p.$ Finally, we point out that our NLL-BFKL predictions are parameter free.

\section{Results for Mueller-Navelet jet $\Delta\Phi$ distributions}

\begin{figure}[t]
\begin{center}
\epsfig{file=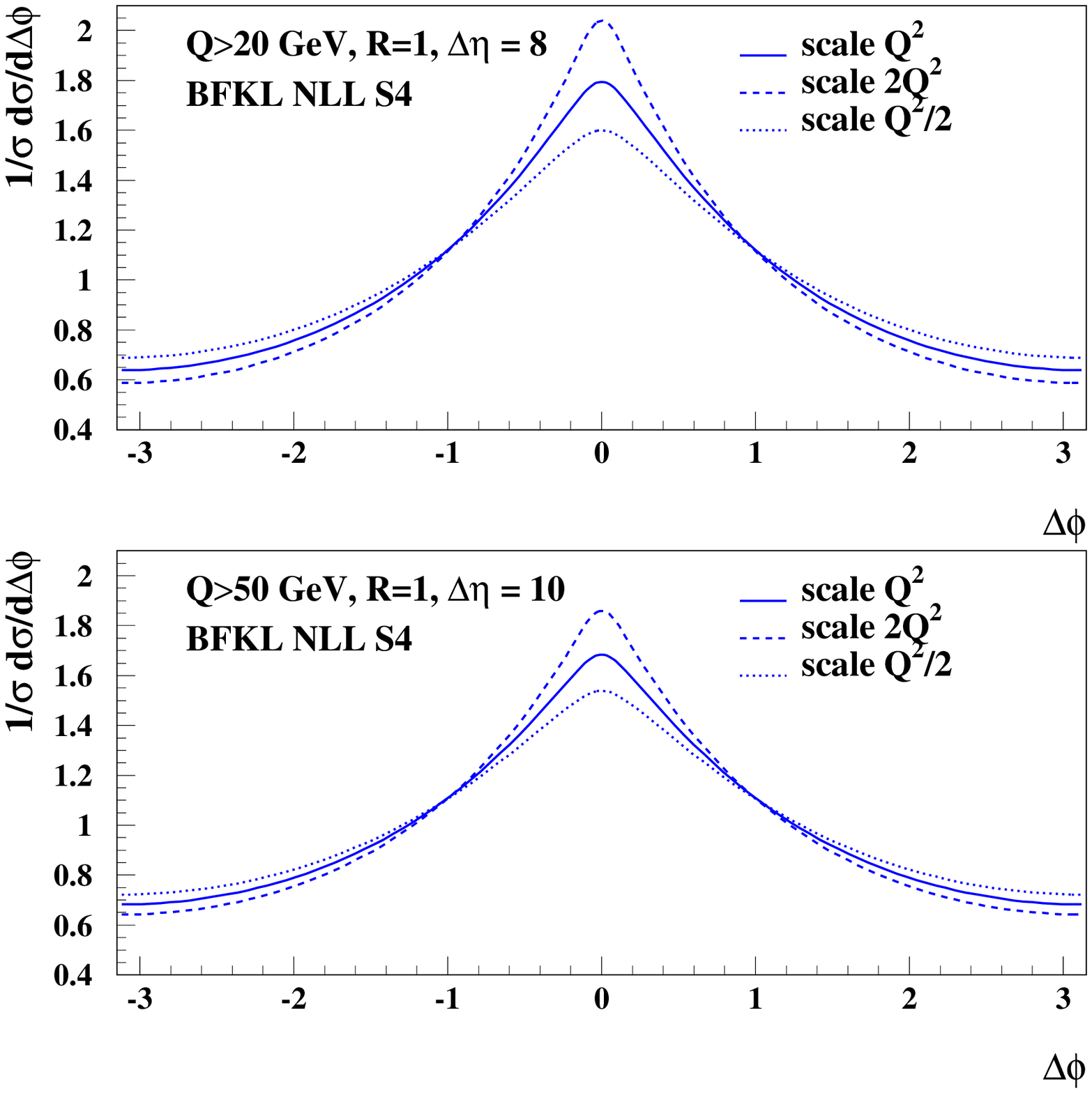,width=4.5cm}
\hspace{1cm}
\epsfig{file=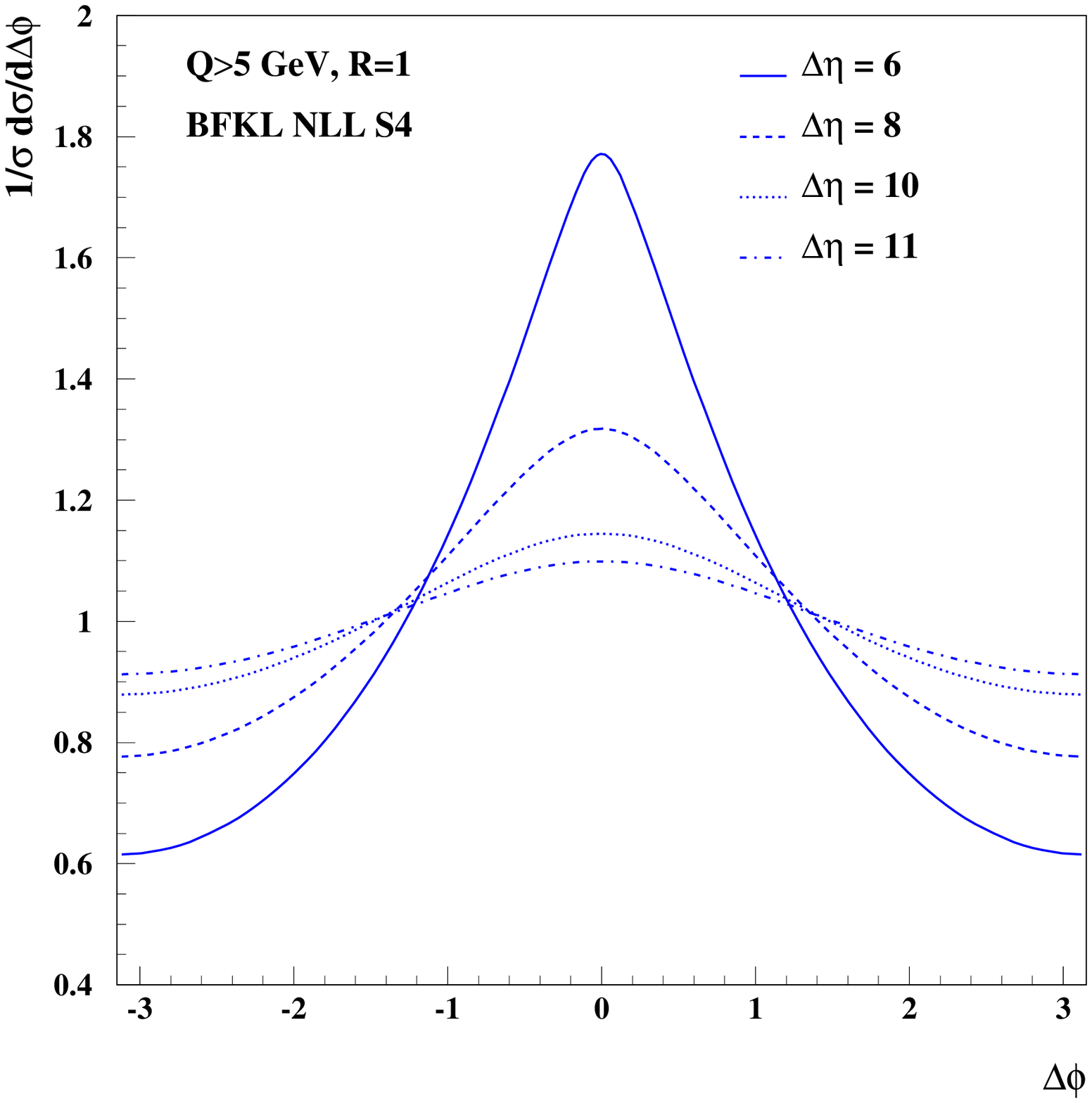,width=4.5cm}
\caption{Left plots: the Mueller-Navelet jet $\Delta\Phi$ spectrum (\ref{obs}) for Tevatron (top) and LHC (bottom) kinematics in the NLL-BFKL framework; the renormalization scale uncertainty is indicated. Right plot: prediction for CDF (miniplugs) kinematics.}
\end{center}
\end{figure}

In general, the $\Delta\Phi$ distribution (\ref{obs}) is peaked around $\Delta\Phi\!=\!0,$ which is indicative of jet emissions occuring back-to-back. In addition the $\Delta\Phi$ distribution flattens with increasing $\Delta\eta$ or with $R$ deviating from 1. In the left part of Fig.1, we display the observable (\ref{obs}) as a function of $\Delta\Phi$, for Tevatron (top) and LHC (bottom) kinematics respectively, and we test the sensitivity of our results when using 
$Q^2/2,\ Q^2$ or $2Q^2$ as the renormalization scale. The scale dependence turns out to be quite small, of about 5 percent, except for $\Delta\Phi$ close to 0, in which case the uncertainty reaches 20 percent.

On the right plot, predictions for the $\Delta\Phi$ distribution with CDF kinematics are presented. Due to miniplug detectors in the forward and backward regions which allow to increase the acceptance in rapidity and transverse momentum to measure very forward jets, it will indeed be possible to measure jets separated in rapidity by more than 10 units and with transverse momenta as low as $5\ \mbox{GeV}.$ With such low values of transverse momenta and large values of rapidity interval between the two jets, it is also likely that saturation effects could play an important role. First estimations \cite{mnjsat} (with less favorable kinematics) indicate so when considering saturation effets damping the LL-BFKL exponential growth. The implementation of saturation effects with the NLL-BFKL growth certainly deserves to be studied.

\end{document}